# Wigner molecules in carbon-nanotube quantum dots

Andrea Secchi[1,2] and Massimo Rontani[1, *]

[1]*S3, Istituto di Nanoscienze – CNR, Via Campi 213/A, 41125 Modena, Italy*
[2]*Dipartimento di Fisica, Università degli Studi di Modena e Reggio Emilia, Via Campi 213/A, 41125 Modena, Italy*

(Dated: June 9, 2010)

We demonstrate that electrons in quantum dots defined by electrostatic gates in semiconductor nanotubes freeze orderly in space realizing a 'Wigner molecule'. Our exact diagonalisation calculations uncover the features of the electron molecule, which may be accessed by tunneling spectroscopy —indeed some of them have already been observed by Deshpande and Bockrath [Nature Phys. **4**, 314 (2008)]. We show that numerical results are satisfactorily reproduced by a simple *ansatz* vibrational wave function: electrons have localized wave functions, like nuclei in an ordinary molecule, whereas low-energy excitations are collective vibrations of electrons around their equilibrium positions.

## I. INTRODUCTION

The paradigm of few-electron complexes in quantum dots (QDs) relies on the 'particle-in-a-box' idea that lowest-energy orbitals are filled according to Pauli's exclusion principle.[1,2] If Coulomb repulsion is sufficiently strong to overcome the kinetic energy cost of localization, a different scenario is predicted:[2,3] a 'Wigner' molecule (WM) forms, made of electrons frozen in space according to a geometrical pattern. Despite considerable experimental effort,[4] evidence of the WM in semiconductor QDs has been elusive so far. In this Article we demonstrate theoretically that WMs occur in gate-defined QDs embedded in typical semiconducting carbon nanotubes (CNTs). Their signatures must be searched —and indeed some of them have already been observed[5]— in tunneling spectra. Through exact diagonalisation (ED) calculations,[6] we unveil the inherent features of the electron molecular states. We show that, like nuclei in a usual molecule, electrons have localized wave functions and hence negligible exchange interactions. The molecular excitations are vibrations around the equilibrium positions of electrons. ED results are well reproduced by an *ansatz* vibrational wave function, which provides a simple theoretical model for chemical potentials and charging energies of ultraclean CNTs.[5,7–9]

In graphene —the unrolled CNT— the ratio of Coulomb potential to kinetic energy is too small to expect Wigner crystallization as well as it is unaffected by carrier density, the usual tuning parameter.[10] On the other hand, the kinetic energy $\sim \hbar\omega_0$ associated to the confinement into a dot embedded in a semiconducting CNT may be controlled by an external gate.[7–9] By keeping the electron number $N$ fixed and decreasing $\hbar\omega_0$, one decreases the density as well to enforce the WM state.

The Coulomb-to-kinetic-energy ratio may be expressed in terms of the dimensionless length per electron $r_s$ [$r_s = 1/(na_B^*)$ with $n$ being the electron density and $a_B^*$ the effective Bohr radius]. Remarkably, in QDs embedded in semiconducting nanotubes $r_s$ is typically one order of magnitude larger than the analogous two-dimensional quantity for nanostructured semiconductor QDs. For example, $r_s \approx 40$ for the CNT dot of Fig. 2(a) whereas $r_s \approx 2$ for the GaAs quasi two dimensional dots studied in Refs. 4. Therefore CNTs are excellent solid-state candidates for the realization of the WM state.

At low energies, electron degrees of freedom in the directions perpendicular to the CNT axis $y$ are frozen, hence the QD is effectively one-dimensional (1D). Wigner crystallization in such a system is not fully understood yet. Since the long-range order of the 1D crystal is smeared by quantum fluctuations, a possible theory relies on the Luttinger liquid model in the presence of long-range interactions.[11,12] However, semiconducting CNT dots have properties which are not easily included in Luttinger theory, like quadratic dispersion relation[13] and quantum confinement.[14]

We exploit the paradigm of the WM, alternative to the Luttinger model, to interpret the outcome of our EDs. In our numerically exact many-body calculations of both ground and excited states we include intra- and inter-valley Coulomb scattering processes[15–17] as well as spin-orbit coupling.[15,16] The envelope-function parts of single-particle states,[18,19] slowly varying with respect to the lattice constant $a$, are eigenstates of a 1D harmonic oscillator of frequency $\omega_0$, which is the generic low-energy model for gate-induced confinement along the CNT axis.[2,15,17]

With respect to previous WM literature,[2,15–17] in this Article we provide: (i) an effective envelope-function theory for low-energy states of ultraclean CNT dots, including the crucial effects of a large energy gap, valley degeneracy, and spin-orbit interaction; (ii) a criterion for crystallization, which is non-trivial for finite systems, as well as it may be experimentally accessed; (iii) a novel ansatz wave function for both ground and excited states, validated by ED.

The structure of this Article is the following: The first four sections detail the theoretical method. In particular, Sec. II introduces CNT bulk states, Sec. III models the QD, Sec. IV reports the fully-interacting many-body Hamiltonian, and Sec. V explains the exact-diagonalization algorithm. *The hasty reader may* skip these technical sections and *go directly to Sec. VI*, which is self-contained and reports the predicted phase diagram for the WM state. Then Sec. VII shows that the WM is indeed a molecule made of electrons, by means of providing an ansatz wave function which compares well with ED data. After the Conclusion (Sec. VIII), Appendix A details the normalization of single-particle wave functions, Appendix B explains the calculation of two-body matrix elements, Appendix C discusses the effect of spin-orbit interaction on WM energy levels, Appendix D provides WM equilibrium positions and eigenfrequencies of the normal modes of vibration.

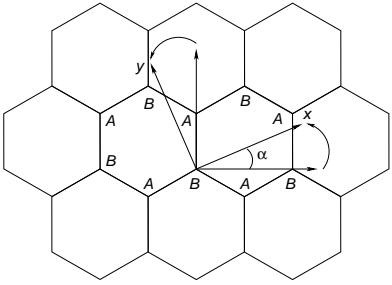

FIG. 1: Schematic representation of the graphene lattice. $\alpha$ is the CNT chiral angle.

## II. NANOTUBE BULK STATES

Consider the two inequivalent triangular sublattices $A$ and $B$ of ideal graphene. The origin of the $xy$ frame is on a $B$-site and the $y$ axis is parallel to the vector connecting the origin to its nearest-neighbor $A$-site (Fig. 1). The lattice sites $\boldsymbol{R}_A$ and $\boldsymbol{R}_B$ are identified by the integers $n_1$ and $n_2$, $\boldsymbol{R}_A \equiv a(n_1 - n_2/2, \sqrt{3}n_2/2 + 1/\sqrt{3})$, $\boldsymbol{R}_B \equiv a(n_1 - n_2/2, \sqrt{3}n_2/2)$, where $a = 2.46$ Å. The two inequivalent corners of the hexagonal first Brillouin zone are $\boldsymbol{K} \equiv (1/3, 1/\sqrt{3})2\pi/a$ and $\boldsymbol{K}' \equiv (2/3, 0)2\pi/a$.

The CNT is obtained by wrapping the graphene sheet along a direction identified by the chiral vector.[20,21] We introduce a rotated reference frame such that the $x$ axis is parallel to the chiral vector and $y$ identifies the CNT axis (see Fig. 1). The new coordinates of atomic positions and wave vectors are obtained by applying the rotation matrix

$$\mathbb{R}(\alpha) = \begin{pmatrix} \cos(\alpha) & \sin(\alpha) \\ -\sin(\alpha) & \cos(\alpha) \end{pmatrix}$$

to old vectors, with $\alpha$ being the chiral angle. Throughout this Article we use the same symbols for new and old coordinates. Hence $x \in [0, 2\pi R[$ is the curvilinear coordinate along the CNT circumference, $y \in [-L_y/2, +L_y/2[$ where $L_y$ is the CNT length, and $z \in [-L_z/2, +L_z/2]$ is orthogonal to the CNT surface. $R$ is the CNT radius and $L_z$ is the characteristic length associated with $2p_z$ orbitals, whose amplitude may be neglected outside the domain of $z$.

Close to the charge neutrality point CNT electrons have energies $E_c(\boldsymbol{K} + \boldsymbol{k}) = \gamma(k_x^2 + k_y^2)^{1/2}$ [$E_c(\boldsymbol{K}' + \boldsymbol{k}') = \gamma(k'^2_x + k'^2_y)^{1/2}$] in valley $\boldsymbol{K}$ ($\boldsymbol{K}'$), where $\boldsymbol{k}$ ($\boldsymbol{k}'$) is the wave vector reckoned from $\boldsymbol{K}$ ($\boldsymbol{K}'$). Here $\gamma$ is the $\pi$-band parameter of graphene[20,22] ($\gamma = 533$ meV·nm) and $k_x(n)$ [$k'_x(n)$] is quantized according to $k_x(n) = (n - \nu/3)/R$ [$k'_x(n) = (n + \nu/3)/R$], with $n$ integer and $\nu = \pm 1$ for semiconducting CNTs, $\nu = 0$ for metallic CNTs.[23] We focus on semiconducting tubes, where the two inequivalent conduction-band minima $\boldsymbol{M}$ and $\boldsymbol{M}'$, slightly displaced from points $\boldsymbol{K}$ and $\boldsymbol{K}'$, correspond to $\boldsymbol{k} \equiv [k_x(0), 0]$ and $\boldsymbol{k}' \equiv [k'_x(0), 0]$, respectively.

The Bloch states at band minima are given by

$$\psi_\tau(\boldsymbol{r}) = \sum_{p=A,B} f_\tau^p \, \varphi_{\tau,p}(\boldsymbol{r}), \qquad (1)$$

where we have introduced the *isospin* index $\tau = +1$ ($\tau = -1$) for point $\boldsymbol{M}$ ($\boldsymbol{M}'$). Here the phase factors are $f_{+1}^A = 1$, $f_{+1}^B = +\nu$, $f_{-1}^A = 1$, $f_{-1}^B = -\nu$, and

$$\varphi_{\tau,p}(\boldsymbol{r}) = e^{-i\nu\tau x/(3R)}\psi_{\tau,p}(\boldsymbol{r}) \qquad (2)$$

is the tight-binding state of the $p$th sublattice.[19,20] The isospin $\tau = \pm 1$, labeling the orbital angular momentum quantum, points to the (anti)clockwise rotation along the circumference coordinate $x$ perpendicular to the CNT axis $y$. The tight-binding state (2) is given by a sum over $2p_z$ atomic orbitals, with:

$$\psi_{\tau,p}(\boldsymbol{r}) = e^{i\theta_\tau^p} \frac{1}{\sqrt{N_c}} \sum_{\{\boldsymbol{R}_p\}} e^{i\boldsymbol{K}_\tau \cdot \boldsymbol{R}_p} \phi_{p_z}(\boldsymbol{r} - \boldsymbol{R}_p). \qquad (3)$$

In Eq. (3) the sum runs on the sites $\boldsymbol{R}_p$ of the $p$th sublattice, $N_c$ is the number of unit cells (one cell contains two carbon atoms), $\boldsymbol{K}_\tau$ stays for $\boldsymbol{K}$ ($\boldsymbol{K}'$) for $\tau = +1$ ($\tau = -1$), $\phi_{p_z}(\boldsymbol{r})$ is the $2p_z$ atomic orbital, $\theta_{+1}^A = 0$, $\theta_{+1}^B = \alpha + 5\pi/3$, $\theta_{-1}^A = \alpha$, $\theta_{-1}^B = 0$. The normalization of $p_z$ orbitals is:

$$\int_{\mathrm{CNT}} |\phi_{p_z}(\boldsymbol{r} - \boldsymbol{R}_p)|^2 \, d\boldsymbol{r} = \mathcal{V}_{\mathrm{CNT}}, \qquad (4)$$

where the integration is over the whole CNT and $\mathcal{V}_{\mathrm{CNT}} = L_x L_y L_z$.

## III. QUANTUM-DOT SINGLE-PARTICLE STATES

We consider a QD embedded in a semiconducting CNT whose length scale, $\ell_{\mathrm{QD}}$ —typically of the order of 10 nm— is much smaller than the CNT length —of order $10^2$–$10^3$ nm, $\ell_{\mathrm{QD}} \ll L_y$.[5,7] Since $\ell_{QD}$ is the relevant single-particle (SP) length the effects of CNT boundaries may be neglected and $k_y$ considered as quasi-continuous, as opposed to the level quantization due to the CNT finite size.[24,25] We assume the quantum confinement to be induced by a gate-modulated soft potential,[5,7] and consider as generic functional form the quadratic potential

$$\frac{1}{2}m^*\omega_0^2 y^2,$$

where $m^* = \hbar^2/(3R\gamma)$ is the effective mass,[18,19] $\omega_0$ is a characteristic frequency, and $\ell_{\text{QD}} = (\hbar/m^*\omega_0)^{1/2}$. In fact, for a soft potential, the first term of its series expansion is quadratic.[26]

Since $\ell_{\text{QD}} \gg a$, SP dot wave functions

$$\psi_{n\tau}(\boldsymbol{r}) = \mathcal{N} F_n(y) \psi_\tau(\boldsymbol{r}) \tag{5}$$

are written[15,18,19] as products of Bloch states $\psi_\tau(\boldsymbol{r})$, rapidly oscillating on the length scale $a$, multiplied by the slowly-varying envelope functions $F_n(y)$, eigenstates of the one-dimensional harmonic oscillator ($n = 0, 1, 2, \ldots$). In Eq. (5) the normalization constant $\mathcal{N}$ is such that

$$\int_{\text{CNT}} \psi^*_{n\tau}(\boldsymbol{r}) \psi_{n'\tau'}(\boldsymbol{r}) \, d\boldsymbol{r} = \delta_{n,n'}\delta_{\tau,\tau'}, \tag{6}$$

with

$$\int_{-\infty}^{+\infty} F_n^*(y) F_{n'}(y) dy = \ell_{\text{QD}} \delta_{n,n'}. \tag{7}$$

In Appendix A we show that the normalization factor is

$$\mathcal{N} = \frac{1}{(2L_x L_z \ell_{\text{QD}})^{1/2}}, \tag{8}$$

where $L_x = 2\pi R$.

The single-particle energy $\varepsilon_{n\tau\sigma}$ is the sum of four contributions:

$$\varepsilon_{n\tau\sigma} = \frac{\gamma}{3R} + \hbar\omega_0\left(n + \frac{1}{2}\right) + \nu\Delta_{\text{SO}}\frac{\gamma}{R}\tau\sigma$$
$$+ \mu_B B\left(\frac{g^*}{2}\sigma - \nu\frac{mR\gamma}{\hbar^2}\tau\right). \tag{9}$$

The first term is half the CNT energy gap, i.e., the distance between the bottom of conduction band and the point in the middle of the gap, taken as a reference. The second one is the oscillator energy. The third contribution accounts for spin-orbit coupling, entangling spin and isospin parts of the wave function (the spin projection along $y$ is $\sigma = \pm 1$). Here we consider only the dominant effect of CNT curvature,[7,15,16,27,28] taking as dimensionless coupling constant $\Delta_{\text{SO}} = 1.25 \cdot 10^{-3}$. The last addendum is the Zeeman term coupling the (iso)spin magnetic dipole with the magnetic field $B$ applied along $y$. Here $\mu_B$ is the Bohr magneton, $g^* = 2$ is the effective giromagnetic factor, $m$ is the free electron mass.

## IV. MANY-BODY HAMILTONIAN

The QD many-body Hamiltonian $\hat{H}$ is the sum of one- and two-body operators, expanded on the basis of single-particle states $\psi_{n\tau}(\boldsymbol{r})$ discussed above:

$$\hat{H} = \hat{H}_{\text{SP}} + \hat{V}_{\text{FW}} + \hat{V}_{\text{BW}}. \tag{10}$$

The single-particle term $\hat{H}_{\text{SP}}$ takes into account the orbital filling,

$$\hat{H}_{\text{SP}} = \sum_{n\tau\sigma} \varepsilon_{n\tau\sigma} \hat{c}^\dagger_{n\tau\sigma} \hat{c}_{n\tau\sigma}, \tag{11}$$

where $\hat{c}^\dagger_{n\tau\sigma}$ creates an electron with spin $\sigma$ in the orbital state $\psi_{n\tau}(\boldsymbol{r})$. The forward (FW)

$$\hat{V}_{\text{FW}} = \frac{1}{2} \sum_{nmpq} \sum_{\tau\tau'} \sum_{\sigma\sigma'} V_{n\tau,m\tau';p\tau',q\tau} \hat{c}^\dagger_{n\tau\sigma} \hat{c}^\dagger_{m\tau'\sigma'} \hat{c}_{p\tau'\sigma'} \hat{c}_{q\tau\sigma} \tag{12}$$

and backward (BW)

$$\hat{V}_{\text{BW}} = \frac{1}{2} \sum_{nmpq} \sum_{\tau\neq\tau'} \sum_{\sigma\sigma'} V_{n\tau,m\tau';p\tau,q\tau'} \hat{c}^\dagger_{n\tau\sigma} \hat{c}^\dagger_{m\tau'\sigma'} \hat{c}_{p\tau'\sigma'} \hat{c}_{q\tau\sigma} \tag{13}$$

two-body operators rule Coulomb scattering processes, with electrons respectively conserving and exchanging their valley location in the reciprocal space.[21] Both FW and BW processes conserve the total lattice momentum. Other scattering channels which do not conserve momentum have been neglected since they are order of magnitudes smaller.

The Coulomb matrix element is written as

$$V_{n\tau,m\tau';p\tau'',q\tau'''} = \ell_{\text{QD}}^{-2} \iint F_n^*(y) F_m^*(y')$$
$$\times\; U_{\tau,\tau';\tau'',\tau'''}(y,y') F_p(y') F_q(y) \, dy\, dy', \tag{14}$$

where $U_{\tau,\tau';\tau'',\tau'''}(y,y')$ is an effective 1D interaction potential, given by

$$U_{\tau,\tau';\tau'',\tau'''}(y,y') = \frac{1}{4L_x^2 L_z^2} \iint \psi^*_\tau(\boldsymbol{r}) \psi^*_{\tau'}(\boldsymbol{r}') U(\boldsymbol{r}-\boldsymbol{r}')$$
$$\times\; \psi_{\tau''}(\boldsymbol{r}') \psi_{\tau'''}(\boldsymbol{r}) \, dr_\perp^{(2)} dr'^{(2)}_\perp, \tag{15}$$

with the integration performed over the coordinates perpendicular to $y$. Here the two-body potential

$$U(\boldsymbol{r}-\boldsymbol{r}') = U_0[1 + \epsilon^2 |\boldsymbol{r}-\boldsymbol{r}'|^2 U_0^2/e^4]^{-1/2} \tag{16}$$

is the Ohno potential, interpolating the two limits of Coulomb-like long-range and Hubbard-like short-range interactions —the latter limit is the Hartree contribution of the $2p_z$ orbital of a single carbon site.[15,21,29,30] In Eq. (16) $U_0 = 15$ eV and $\epsilon$ is the CNT dielectric constant.

As we show in Appendix B, manipulation of Eq. (15) provides the effective 1D potential for FW scattering $U_{\text{FW}}(y-y') \equiv U_{\tau,\tau';\tau',\tau}(y-y')$:

$$U_{\text{FW}}(y-y') = \frac{2e^2}{\pi\epsilon} \frac{K\left(2R[4R^2+\mathcal{L}^2+a_z^2+(y-y')^2]^{-1/2}\right)}{[4R^2+\mathcal{L}^2+a_z^2+(y-y')^2]^{1/2}}, \tag{17}$$

where $\mathcal{L} = e^2/(U_0\epsilon)$, $K(k)$ is the complete elliptic integral of the first kind ($0 \leq k^2 < 1$),[31] and $a_z = 3a_B$, with $a_B$ being the Bohr radius. Note that, as $|y-y'| \to \infty$ in Eq. (17), $K \to \pi/2$ and $U_{\text{FW}}(y-y')$ behaves as the Coulomb potential. On the other hand, BW scattering is short-ranged with respect to the length scale $\ell_{\text{QD}}$, with typical matrix elements being orders of magnitude smaller than those for FW scattering. Explicitly, the effective 1D BW potential $U_{\text{BW}}(y-y') \equiv U_{\tau,-\tau;\tau,-\tau}(y-y')$ is:

$$U_{\text{BW}}(y-y') = \tilde{U}_{\text{BW}} \frac{\sqrt{3}a^2}{8\pi R} \delta(y-y'), \qquad (18)$$

where $\tilde{U}_{\text{BW}} \approx 4$ eV is a constant whose value is determined by the lattice structure.[21]

By using expressions (17) and (18) into (14) we are left with a two-dimensional numerical quadrature involving the envelope functions $F_n(y)$. Note that $\hat{V}_{\text{FW}}$ conserves the total (iso)spin $S$ ($T$) as well as its projection $S_y$ ($T_y$), whereas the symmetry-breaking effect of $\hat{V}_{\text{BW}}$ is negligible. The spin-orbit term appearing in $\hat{H}_{\text{SP}}$ conserves only $S_y$ and $T_y$.

## V. EXACT DIAGONALIZATION

We solve the $N$-body problem by exactly diagonalizing $\hat{H}$ [Eq. (10)], which is a matrix in the Fock space of Slater determinants $|\Phi_i^N\rangle$ (the method is also known as full configuration interaction).[6] We build the Slater determinants $|\Phi_i^N\rangle$ by filling in all possible ways with $N$ electrons the $N_{\text{SP}}$ lowest-energy SP orbitals $\psi_{n\tau}(\boldsymbol{r})$, twofold spin-degenerate when $B = 0$ and $\Delta_{\text{SO}} = 0$. We take $N_{\text{SP}} = 30$ for the extensive ground-state calculations of Figs. 2 and 4 and $N_{\text{SP}} = 50$ otherwise. Both ground and excited many-body states $\left|\Psi_N^{(n)}\right\rangle$, written as linear combinations of Slater determinants,

$$\left|\Psi_N^{(n)}\right\rangle = \sum_i c_i^{(n)} |\Phi_i^N\rangle, \qquad (19)$$

are obtained numerically, together with their energies, by means of the parallel code *DonRodrigo*.[32] The code output [i.e., the coefficients $c_i^{(n)}$] is then post-processed in order to calculate the charge density $n(y)$ and pair correlation function $g(y)$ for a given state (see Sec. VII). The diagonalization proceeds in each Hilbert space sector labeled by $N$, $S_y$, and the parity of the total envelope wave function under spatial reflection $y \to -y$. Note that the symmetry-breaking effect of spin-orbit interaction largely increases sector matrix sizes by mixing blocks labeled by different values of $S$ (the maximum linear size we have managed is $883,232$ for $N = 5$ and $N_{\text{SP}} = 30$). The relative error for low-lying excitation energies, estimated for the Kohn (center-of-mass) mode with $N = 4$, is smaller than $10^{-7}$.

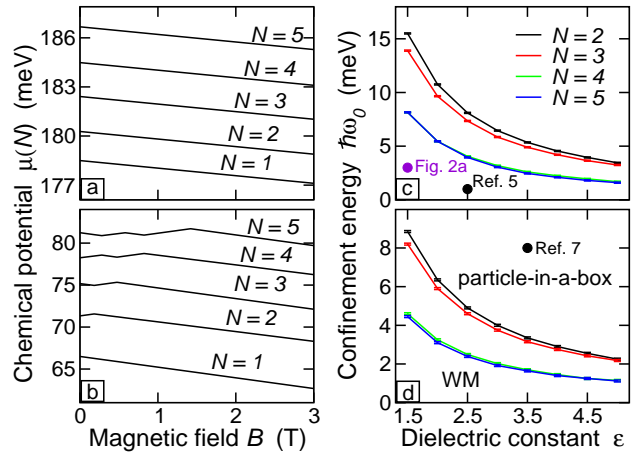

FIG. 2: (color online). (a) Typical WM chemical potentials $\mu(N)$ vs $B$ for $1 \leq N \leq 5$ electrons. The curves for $N = 2, 3, 4, 5$, were rigidly shifted by -20, -38, -55, -72 meV, respectively. Whereas at $B = 0$ T ground states are highly degenerate (within an energy range of at most $\approx 1$ $\mu$eV), the field selects the (iso)spin-polarized states. (b) Typical particle-in-a-box $\mu(N)$ vs $B$ for $R = 3$ nm, $\hbar\omega_0 = 15$ meV, $\epsilon = 3$. The curves for $N = 2, 3, 4, 5$, were rigidly shifted by -27, -52, -75, -97 meV, respectively. (c) WM phase diagram in the $(\epsilon, \hbar\omega_0)$ space for $R = 1$ nm and $2 \leq N \leq 5$. The vertical error bar of each point corresponds to 0.1 meV. The violet [gray] (black) dot identifies the QD of Fig. 2(a) (the device D1 of Ref. 5, with $R = 0.8$ nm). (d) Same as Fig. 2(c) for $R = 3$ nm. The black dot identifies the QD device of Ref. 7 ($R = 3.6$ nm).

## VI. PHASE DIAGRAM

The WM is made of localized electrons whose mutual exchange interactions are negligible,[3,33] so no energy is required to orient all spins $\sigma$ along a magnetic field $B$ parallel to the CNT axis, $y$. Similarly, in the WM state Coulomb interactions between electrons do not depend on their isospins $\tau$ —the orbital angular momentum along $y$ labeling valleys $\boldsymbol{K}$ ($\tau = +1$) and $\boldsymbol{K'}$ ($\tau = -1$) in the reciprocal space.[22] These features, which hold also in the presence of spin-orbit coupling (cf. Appendix C), are fingerprints of the WM and may be directly observed from the slopes of the chemical potentials $\mu(N)$ measured in tunneling experiments.[1]

To validate this claim, we obtain $\mu(N) = E_0(N) - E_0(N-1)$ through the ED calculation of ground state energies $E_0(N)$ for consecutive electron numbers. Figures 2(a) and (b) show the predicted $\mu(N)$ vs $B$ in the two exemplary cases of WM [Fig. 2(a)] and particle-in-a-box [Fig. 2(b)] ground states, respectively.

In Fig. 2(a), computed for a realistic QD with $\hbar\omega_0 = 3$ meV, CNT radius $R = 1$ nm, dielectric constant $\epsilon = 1.5$, all curves are parallel straight lines pointing downward in energy with $B$. This is distinctive of the WM, since $\mu(N)$ depends on $B$ only through the single-particle (iso)spin Zeeman terms and each added electron enters the QD

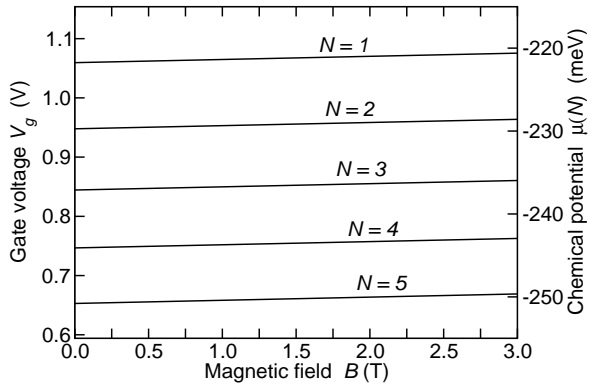

FIG. 3: Typical WM chemical potentials $\mu(N)$ (vertical right axis) vs $B$ for $1 \leq N \leq 5$ holes, for $R = 0.8$ nm, $g^* = 2$, $\hbar\omega_0 = 1$ meV, $\epsilon = 2.5$. The vertical left axis is obtained by converting energies (vertical right axis) into voltages by using the conversion factor ($= 14$) estimated in Ref. 5 with an arbitrary shift.

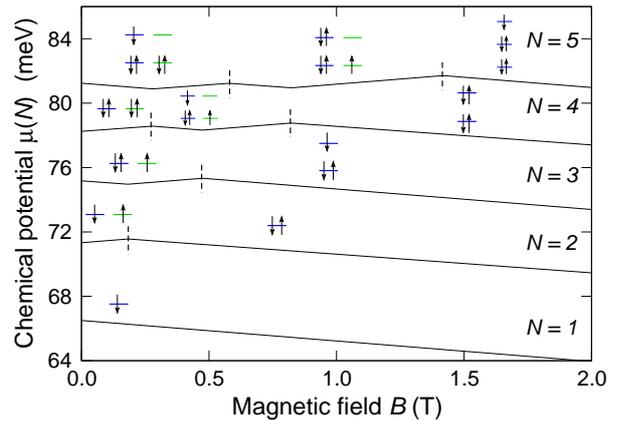

FIG. 4: (color online). Filling sequence for the spectrum reported in Fig. 2(b). The insets point to the Slater determinants with the largest weights in the ED expansions of $N$-body ground states in certain ranges of $B$ (separated by vertical dashed lines). The blue [dark gray] (green [light gray]) ladders of levels depict the lowest harmonic oscillator states for $\tau = +1$ ($\tau = -1$), whereas arrows represent electron spins. Note that the sign of the slope of $\mu(N)$ depends on the sign of the isospin $\tau = \mp1$ of the tunnelling electron injected into the dot already filled by $N - 1$ particles. For $N \geq 2$ one may discriminate between two distinct particle-in-a-box regions along the $B$ axis: (a) one unpolarized phase close to the origin; (b) one isospin polarized, antiferromagnetic phase at finite values of $B$ (the larger $N$, the stronger $B$). Regions (a) and (b) map respectively into the experimental phases III and II reported in Ref. 5. For larger values of $B$ the spin- and isospin-polarized WM state (experimental phase I) is recovered.

with the same (iso)spin aligned to $B$ —to minimize the magnetic dipole energy.

Remarkably, spectra like those of Fig. 2(a) were recently observed up to $N \sim 10$ holes in ultraclean QDs embedded in gated suspended tubes.[5] Indeed, the ED hole spectrum reported in Fig. 3 compares well with Fig. 2c of Ref. 5 (phase I). In particular, the energy separation of $\approx 8$ meV between adjacent $\mu(N)$ at $B = 0$ agrees well with the experimental data. Note that the parameters of the ED calculation of Fig. 3 have been taken from the estimates given in the main text of Ref. 5 when available ($R$ and $g^*$), whereas the unavailable parameters ($\epsilon$ and $\hbar\omega_0$) were obtained by fitting the plots of $\mu(N)$ vs $B$, for different $N$, to the experimental curves. Importantly, the number of free parameters in our calculation (two) is much smaller than the number of experimental constraints [five $\mu(N)$ curves]. Besides, it is worth noting that the gapped Luttinger liquid theory[12] used to explain the phase diagram of Ref. 5 accounts for neither the large energy gap ($\sim 220$ meV) nor spin-orbit coupling.[7]

The particle-in-a-box model is recovered by significantly reducing the Coulomb-to-kinetic energy ratio, which is accomplished by increasing $\hbar\omega_0$, $R$, or $\epsilon$ (the latter is strongly sensitive to the presence of external leads), as shown in Fig. 2(b). Close to $B = 0$ T the curves have now slopes depending on $N$, according to the values of $\sigma$ and $\tau$ of tunneling electrons. The latter manifest the filling of the QD orbitals and are ruled by Hartree, exchange, and spin-orbit interactions[2,22,34] (the shell filling sequence is shown in Fig. 4). Whereas for finite values of $B$ the $\mu(N)$ of Fig. 2(b) show kinks due to the crossings of competing ground states, the slopes of Fig. 2(a) are perfectly constant and negative. This disparate behavior may be easily discerned experimentally, providing an operative definition of the WM phase.

The WM phase diagram in the $(\epsilon, \hbar\omega_0)$ space is shown in Fig. 2(c) [Fig. 2(d)] for $R = 1$ nm (3 nm). The WM region identifies the locus of parameters for which $\mu(N)$ is a straight line along the $B$-axis pointing downward. Since no sharp phase transition occurs to the WM, boundaries depend on $N$ —an effect of the finite size of the system. Specifically, the position of the boundary line depends on the excitation energy of the lowest state with maximum (iso)spin. This energy is almost constant for $N = 2, 3$, related to a change of the orbital parity. A second contribution adds for $N = 4, 5$, due to additional (iso)spin flips. As $N$ increases, a smaller value of either $\epsilon$ or $\hbar\omega_0$ is required to enter the WM phase [for fixed $\hbar\omega_0$ the density increases with $N$ (Ref. 2)]. Hence boundary lines accumulate in the bottom regions of the plots.

The black dots shown in Figs. 2(c) and (d) point to the QD devices studied in Refs. 5 and 7, respectively. In both cases the computed plots of $\mu(N)$ vs $B$ nicely match their experimental counterparts (cf. Fig. 3 and Ref. 15). Overall, in Figs. 2(c) and (d) both the WM and particle-in-a-box phases are ground states for a broad range of values of $\hbar\omega_0$ and $\epsilon$ which may be realistically achieved in current experiments.[5,7–9]

Figure 2(c) suggests that, for the QD of Ref. 5 (black dot), one exits the WM phase by increasing $N$ (boundary lines move at lower values of $\hbar\omega_0$). In Ref. 5, the WM state (there labeled phase I) is replaced by the isospin polarized, spin antiferromagnetic phase II at $N \sim 10$, and then by the unpolarized phase III at $N \sim 15$. This scenario is consistent with the phase diagrams of Fig. 2. In fact, as illustrated in Fig. 4 for $N \geq 2$, phases II and III map into different particle-in-a-box regions along the $B$-axis: the latter is unpolarized close to $B = 0$, the former is isospin polarized at finite $B$. Note that strong values of $B$ eventually induce phase I, i.e., the WM state. By moving upwards along the $\hbar\omega_0$-axis in the diagram of Fig. 2(c), we expect the critical values of $N$ to decrease as well as the region II of the particle-in-a-box phase to vanish close to $B = 0$.

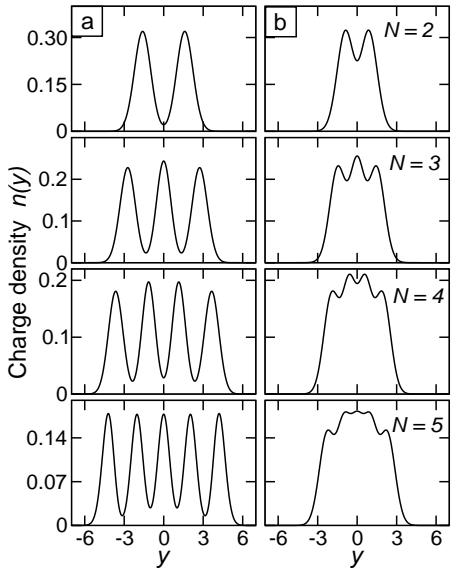

FIG. 5: Charge density $n(y)$ vs $y$ for $2 \leq N \leq 5$ for the ground state at $B = 0$ T. The length unit is the QD length $\ell_{\mathrm{QD}} = (\hbar/m^*\omega_0)^{1/2}$. (a) The parameters used for the ED calculations were $R = 1$ nm, $\hbar\omega_0 = 3$ meV, $\epsilon = 1.5$, as in Fig. 2(a) ($\ell_{\mathrm{QD}} = 23$ nm). (b) The parameters used were $R = 3$ nm, $\hbar\omega_0 = 15$ meV, $\epsilon = 3$, as in Fig. 2(b) ($\ell_{\mathrm{QD}} = 18$ nm).

## VII. WIGNER MOLECULE ANSATZ WAVE FUNCTION

In the following we provide direct evidence that WM states are indeed molecules made of electrons. To this aim, we first plot in Fig. 5 the envelope-function part of the charge density

$$n(y) = \frac{1}{N}\sum_{i=1}^{N}\langle\delta(y - y_i)\rangle$$

at zero field, for the same two sets of parameters as in Figs. 2(a) and (b) [Figs. 5(a) and (b), respectively]. Here $y_i$ is the coordinate of the $i$th electron and $\langle\ldots\rangle$ is the quantum average for a certain state. For the WM ground state with $N$ electrons, $n(y)$ displays $N$ clearly resolved peaks of approximately equal weights [Fig. 5(a)], whereas in Fig. 5(b) it shows a compact droplet with a faint structure superimposed. The charge inhomogeneity of Fig. 5(a) is due to the spatial localization of electrons, controlled by the competing effects of Coulomb repulsion and confinement of the harmonic potential [the length unit $\ell_{\mathrm{QD}} = (\hbar/m^*\omega_0)^{1/2}$ is the characteristic length of the oscillator]. Whereas in Fig. 5(a) Coulomb interaction breaks the spatial homogeneity of the electron droplet, in Fig. 5(b) the effect of confinement is preponderant, squeezing the charge density.

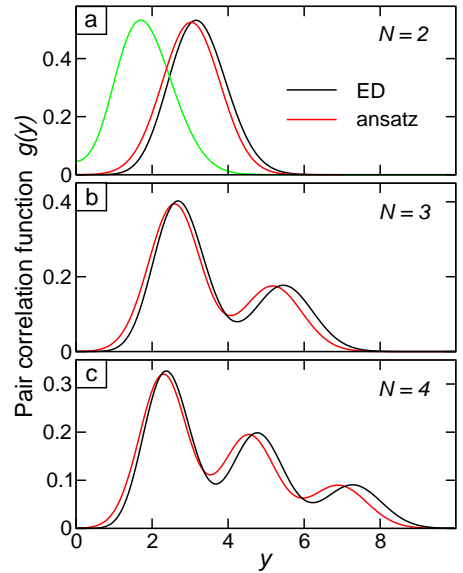

FIG. 6: (color online). WM pair correlation function $g(y)$ vs $y$ for the ED (black curves) and ansatz (red [dark gray] curves) ground states. The length unit is $\ell_{\mathrm{QD}}$ and $\int_0^\infty g(y)dy = 1$. The parameters used are the same as in Figs. 2(a) and 5(a). (a) Two-electron ground state. The green [light gray] curve shows the particle-in-a-box result [ED parameters as in Figs. 2(b) and 5(b)]. (b) Three-electron ground state. (c) Four-electron ground state.

We note that the density in the valleys between two consecutive peaks in the plots of Fig. 5(a) is finite. This behavior is consistent with the fact that exchange interactions in the WM are negligible. In fact, to assess the wave function overlap between two electrons one needs to compute the *two-body* correlation function

$$g(y) \propto \sum_{i\neq j}\langle\delta(y - y_i + y_j)\rangle,$$

giving the probability of finding two electrons at relative position $y$, whereas $n(y)$ is a one-body observable. In Fig. 6(a) we compare the WM with the particle-in-a-box ground states, by plotting $g(y)$ for $N = 2$ (black and green [light gray] curves, respectively). The most remarkable difference is that the probability of being in contact for two electrons is negligible only for the WM [$g(y = 0) \approx 0$ for the black curve]. Hence the mutual (iso)spin orientation is irrelevant to the WM energy, consistently with our criterion for the phase boundary (Fig. 2). Besides, $g(y)$ shows a clearly resolved peak at $y \approx 3$ [black curve of Fig. 6(a)], pointing to the freezing of relative motion around a fixed equilibrium distance.

TABLE I: Selected locations of the peaks of the ED charge density $n(y)$ shown in Fig. 5(a) vs classical equilibrium positions. The length unit is $\ell_{\rm QD}$. Analytical expressions for classical values are provided in Appendix D. The parameters are $\epsilon = 1.5$, $R = 1$ nm, $\hbar\omega_0 = 3$ meV.

| N | location | ansatz | ED |
|---|----------|--------|-----|
| 2 | $\bar{y}_2$ | 1.5 | 1.6 |
| 3 | $\bar{y}_3$ | 2.6 | 2.7 |
| 4 | $\bar{y}_3$ | 1.1 | 1.1 |
| 4 | $\bar{y}_4$ | 3.5 | 3.6 |

In order to build a simple ansatz for WM wave functions, we parallel the construction of the vibrational wave function of poly-atomic molecules.[35] Therefore, we consider $N$ point-like classical particles in the 1D quadratic trap of frequency $\omega_0$, interacting via the Coulomb potential $e^2/\epsilon |y_1 - y_2|$. For the small, harmonic oscillations around equilibrium positions $\bar{y}_i$ we find the normal modes of vibration, with normal coordinates $Y_i$ and eigenfrequencies $\omega_i$, $i = 1, \ldots, N$ (see Appendix D). Then, we quantize the system and write the wave function $\Psi_{\rm vib}$ as

$$\Psi_{\rm vib} = \prod_{i=1}^{N} \Psi_{n_i}(Y_i), \quad (20)$$

where $\Psi_{n_i}(Y_i)$ is the $n_i$th excited state of the harmonic oscillator for the $i$th normal mode of vibration, whose energy is $\hbar\omega_i(n_i + 1/2)$ ($n_i = 0, 1, 2, \ldots$). The total ansatz wave function, $\Psi_{\rm ansatz}(y_1, \tau_1, \sigma_1; \ldots; y_N, \tau_N, \sigma_N)$, is given by the product

$$\Psi_{\rm ansatz} = \mathcal{A}\, \Psi_{\rm vib}\, \Psi_{\rm iso}\, \Psi_{\rm spin} \quad (21)$$

respectively of the vibrational $\Psi_{\rm vib}(Y_1, \ldots, Y_N)$, the isospin $\Psi_{\rm iso}(\tau_1, \ldots, \tau_N)$, and the spin $\Psi_{\rm spin}(\sigma_1, \ldots, \sigma_N)$ parts, where $\mathcal{A}$ is the anti-symmetrization operator and the $Y_i$'s are expressed in terms of the original coordinates $y_i$'s.

To compare the ED and ansatz WM wave functions, we first notice that the locations of the maxima of the ED charge densities $n(y)$ of Fig. 5(a) nicely match the classical equilibrium positions $\bar{y}_i$, as shown in Table I. We then plot the correlation functions $g(y)$ for the ground states —up to four electrons— in Fig. 6. The excellent matching between ED (black curves) and ansatz (red [dark gray] curves) data points to the intrinsic vibrational structure of the WM wave function: In fact, the peaks appearing in $g(y)$ identify the equilibrium distances of localized electrons, whereas the widths of peaks originate from the zero-point motions of oscillators.

The vibrational ansatz is especially useful to obtain both addition and excitation energies of the many-body system, which may be accessed respectively by linear and non-linear tunneling spectroscopy.[1,34] The addition energy —the energy spacing $\Delta\mu(N) = \mu(N+1) - \mu(N)$ between consecutive chemical potentials— is the charging energy required to add one electron to the $N$-body system at zero field.[34] $\Delta\mu(N)$ may be simply calculated within the ansatz vibrational model from the static energies of charges located at equilibrium positions plus their zero-point oscillations. These in turn may be simply worked out from Appendix D. Figure 7 shows the dependence of $\Delta\mu(N)$ on $N$. The agreement between ED data (black circles in Fig. 7) and those predicted from the vibrational ansatz (red [gray] squares) is excellent for all values of $R$, demonstrating the accuracy of the ansatz wave function.

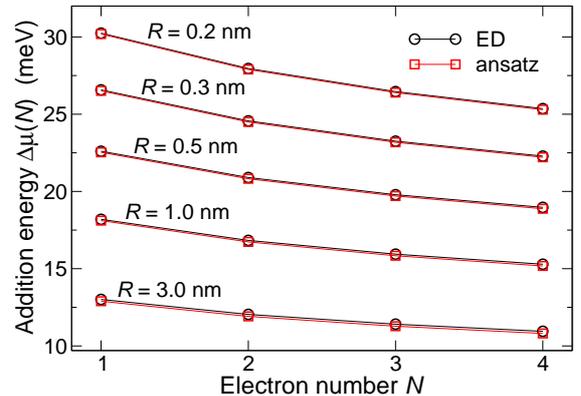

FIG. 7: (color online). Addition energy $\Delta\mu(N)$ vs $N$ for different CNT radii $R$. Black circles (red [gray] squares) are obtained from ED calculations (ansatz vibrational model). The lines are guides to the eye. Here $\epsilon = 2$ and $\hbar\omega_0 = 3$ meV.

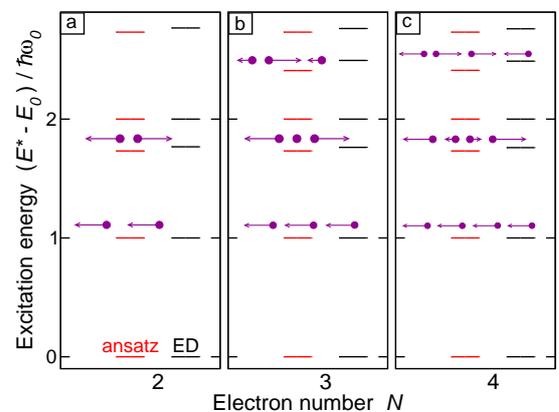

FIG. 8: (color online). Normalized WM excitation energies $(E^* - E_0)/\hbar\omega_0$ according to ED (black lines) and ansatz (red [gray] lines) predictions, in the absence of spin-orbit interaction at $B = 0$ T. The schematic diagrams highlight the classical normal modes of vibrations corresponding to one excitation quantum. The parameters are the same as in Figs. 2(a), 5(a), 6. (a) $N = 2$. (b) $N = 3$. (c) $N = 4$.

Excitation energies are given by simply specifying the

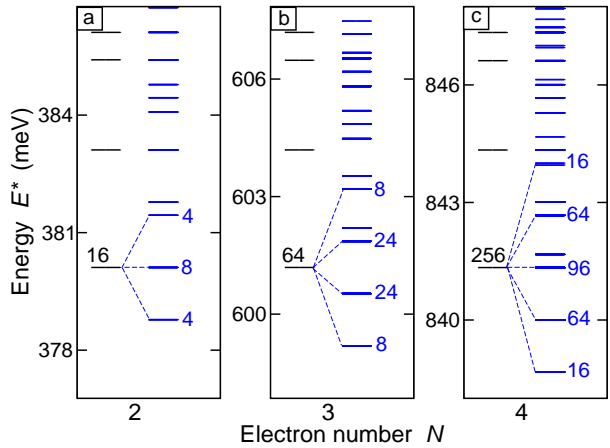

FIG. 9: (color online). ED energies of WM excited states with (blue [gray] lines) and without (black lines) spin-orbit interaction. The numbers point to multiplet degeneracies and the dashed lines connect split multiplets to parent levels. The ED parameters are the same as in Fig. 8. (a) $N = 2$. (b) $N = 3$. (c) $N = 4$.

quanta of the excited vibrational modes:

$$E^*(N) - E_0(N) = \sum_{i=1}^{N} \hbar\omega_i\, n_i, \qquad (22)$$

with $E^*(N)$ being the energy of a certain excited state. To validate this prediction, in Fig. 8 we compare ansatz (red [gray] lines) and ED (black lines) low-lying excitation energies $E^*(N) - E_0(N)$ for $N = 2$ [Fig. 8(a)], 3 [Fig. 8(b)], 4 [Fig. 8(c)]. For the sake of clarity, here we have neglected the effect of spin-orbit interaction. The agreement is very good, particularly at low energies —say less than $\approx 2\hbar\omega_0$— at which both center-of-mass (Kohn) and breathing modes are excited (cf. diagrams of Fig. 8). We attribute the slight discrepancy between black and red [gray] lines of Fig. 8 to the simplified form of the ansatz two-body potential in comparison with the complexity of Coulomb interaction in the CNT.

In the absence of spin-orbit interaction, WM states are highly degenerate. This comes out naturally from the vibrational ansatz, since electrons at their equilibrium positions may freely flip both their spins and isospins in $4^N$ possible ways. This $4^N$-fold degeneracy is confirmed by ED, except for tiny splittings (of few tenths of $\mu$eV) due to the residual electron delocalization. The spin-orbit interaction, induced by the CNT curvature,[7] partially lifts this degeneracy (cf. Fig. 9 and discussion of Appendix C). We checked that both ansatz and ED states have the same (iso)spin structure and hence spin-orbit induced energy splittings. The latter may be easily evaluated analytically (cf. Appendix C).

## VIII. CONCLUSION

In conclusion, WM states occur in CNT dots. Crystallized electrons show a molecular behavior, which is reproduced by an ansatz many-body wave function. The latter provides simple predictions for addition energies and excitation modes of the electron molecule, accessible via tunneling spectroscopy. These findings are relevant for recent transport experiments[8,9] aimed at achieving coherent spin manipulation in ultraclean nanotubes, since they show that the full inclusion of Coulomb correlation is an essential step in the interpretation of chemical potentials and charging energies. Whether the peculiar features of the CNT Wigner molecule, like electron localization, isospin,[36] and spin-orbit coupling,[7,37] may be combined to provide operational protocols for novel quantum devices remains an open issue. We hope this work may help stimulate new experiments along the same path.

### Acknowledgments

This work was supported by INFM-CINECA 2009, MAE Italy-USA 2010 Grant 'Electronic properties of graphene-based nanostructures', and Fondazione Cassa di Risparmio di Modena. We thank M. Grifoni, E. Andrei, S. Ilani, F. Kuemmeth, G. A. Steele, E. Molinari, V. Pellegrini, S. Corni, R. Di Felice, S. Picozzi, F. Troiani for discussions and L. Neri for proofreading the manuscript.

### Appendix A: Normalization of single-particle wave functions

By using Eqs. (1), (2), and (3), we can write the normalization integral (6) as:

$$\begin{aligned}
\int_{\mathrm{CNT}} \psi^*_{n\tau}(\boldsymbol{r})\, \psi_{n'\tau'}(\boldsymbol{r})\, d\boldsymbol{r} &= \frac{|\mathcal{N}|^2}{N_c} \sum_{p,p'} \sum_{\{\boldsymbol{R}_p\}} \sum_{\{\boldsymbol{R}'_{p'}\}} f^p_\tau f^{p'}_{\tau'} \\
&\times\ e^{i\left(\theta^{p'}_{\tau'} - \theta^p_\tau\right)} e^{i\left(\boldsymbol{K}_{\tau'} \cdot \boldsymbol{R}'_{p'} - \boldsymbol{K}_\tau \cdot \boldsymbol{R}_p\right)} \int_{\mathrm{CNT}} F^*_n(y) F_{n'}(y) \\
&\times\ e^{-i\nu(\tau'-\tau)x/(3R)} \phi^*_{p_z}(\boldsymbol{r} - \boldsymbol{R}_p)\, \phi_{p_z}(\boldsymbol{r} - \boldsymbol{R}'_{p'})\, d\boldsymbol{r}.
\end{aligned} \qquad (A1)$$

Since $p_z$ orbitals are strongly localized we neglect two-center overlaps:

$$\phi^*_{p_z}(\boldsymbol{r} - \boldsymbol{R}_p)\, \phi_{p_z}(\boldsymbol{r} - \boldsymbol{R}'_{p'}) \approx |\phi_{p_z}(\boldsymbol{r} - \boldsymbol{R}_p)|^2\, \delta_{p,p'} \delta_{\boldsymbol{R}_p, \boldsymbol{R}'_{p'}}. \qquad (A2)$$

Moreover, with respect to the length scale of envelope functions, atomic orbitals have an almost singular spatial dependence, so we assume[29]

$$|\phi_{p_z}(\boldsymbol{r} - \boldsymbol{R}_p)|^2 \approx \delta(\boldsymbol{r} - \boldsymbol{R}_p)\, \mathcal{V}_{\mathrm{CNT}}, \qquad (A3)$$

consistently with Eq. (4). Inserting Eqs. (A2) and (A3) into (A1) we obtain

$$\int_{\text{CNT}} \psi_{n\tau}^*(\boldsymbol{r})\,\psi_{n'\tau'}(\boldsymbol{r})\,d\boldsymbol{r} = \frac{|\mathcal{N}|^2 \mathcal{V}_{\text{CNT}}}{N_c} \sum_{p=A,B} f_\tau^p f_{\tau'}^p$$
$$\times\ e^{i\left(\theta_{\tau'}^p - \theta_\tau^p\right)} \sum_{\{\boldsymbol{R}_p\}} e^{i(\boldsymbol{K}_{\tau'} - \boldsymbol{K}_\tau)\cdot\boldsymbol{R}_p} F_n^*(R_p^y) F_{n'}(R_p^y). \tag{A4}$$

The last sum in (A4) is the Fourier component of the product $F_n^*(y)F_{n'}(y)$ of wave vector $\boldsymbol{K}_{\tau'} - \boldsymbol{K}_\tau$. The off-diagonal Fourier component ($\tau \neq \tau'$) is negligible with respect to the diagonal one ($\tau = \tau'$) since envelope functions $F_n(y)$ are slowly varying with respect to the lattice scale $a$.[18] By keeping only the leading term all phase factors cancel and

$$\int_{\text{CNT}} \psi_{n\tau}^*(\boldsymbol{r})\,\psi_{n'\tau'}(\boldsymbol{r})\,d\boldsymbol{r}$$
$$= \delta_{\tau,\tau'} \frac{|\mathcal{N}|^2 \mathcal{V}_{\text{CNT}}}{N_c} \sum_{\{\boldsymbol{R}\}} F_n^*(R^y) F_{n'}(R^y), \tag{A5}$$

where $\{\boldsymbol{R}\}$ includes both sublattices. By replacing the sum over lattice sites in (A5) with an integral on the axial coordinate $y$, the following must hold:

$$\sum_{\{\boldsymbol{R}\}} F_n^*(R^y) F_{n'}(R^y) \approx \frac{g(\Delta y)}{\Delta y} \int F_n^*(y) F_{n'}(y) dy, \tag{A6}$$

where $g(\Delta y)/\Delta y$ is the number of atoms occupying a portion of the CNT of length $\Delta y$, given by

$$\frac{g(\Delta y)}{\Delta y} = \frac{2 N_c}{L_y}. \tag{A7}$$

By inserting Eqs. (A7) and (7) into (A6), we find:

$$\sum_{\{\boldsymbol{R}\}} F_n^*(R^y) F_{n'}(R^y) \cong \delta_{n,n'} \frac{2 N_c \ell_{\text{QD}}}{L_y}. \tag{A8}$$

By inserting back (A8) into (A5), one obtains

$$\int_{\text{CNT}} \psi_{n\tau}^*(\boldsymbol{r})\,\psi_{n'\tau'}(\boldsymbol{r})\,d\boldsymbol{r} = \delta_{n,n'}\delta_{\tau,\tau'}\,|\mathcal{N}|^2\,2L_x L_z \ell_{\text{QD}}, \tag{A9}$$

from which Eq. (8) immediately follows.

### Appendix B: Two-body matrix elements

By inserting expansion (1) into (15), using (2) and (3), and then integrating over the coordinates perpendicular to $y$ exploiting (A2) and (A3), one obtains:

$$U_{\tau,\tau';\tau'',\tau'''}(y,y') = \frac{L_y^2}{4 N_c^2} \sum_p f_\tau^p f_{\tau'''}^p e^{i(\theta_{\tau'''}^p - \theta_\tau^p)}$$
$$\times \sum_{p'} f_{\tau'}^{p'} f_{\tau''}^{p'} e^{i(\theta_{\tau''}^{p'} - \theta_{\tau'}^{p'})} \sum_{\{\boldsymbol{R}_p\}} \sum_{\{\boldsymbol{R}'_{p'}\}}$$
$$e^{i\left[(\boldsymbol{M}_{\tau'''} - \boldsymbol{M}_\tau)\cdot\boldsymbol{R}_p + (\boldsymbol{M}_{\tau''} - \boldsymbol{M}_{\tau'})\cdot\boldsymbol{R}'_{p'}\right]}$$
$$\times\ U(\boldsymbol{R}_p - \boldsymbol{R}'_{p'})\,\delta(y - R_p^y)\,\delta(y' - R_{p'}^{\prime y}), \tag{B1}$$

with $\boldsymbol{M}_{+1} = \boldsymbol{M}$ [$\boldsymbol{M}_{-1} = \boldsymbol{M}'$]. FW and BW integrals correspond respectively to the choices of indices $(\tau,\tau';\tau',\tau)$ and $(\tau,-\tau;\tau,-\tau)$ appearing in Eq. (B1).

#### 1. FW integrals

Equation (B1) reads as

$$U_{\text{FW}}(y,y') = \frac{L_y^2}{4 N_c^2} \sum_{\{\boldsymbol{R}\}} \sum_{\{\boldsymbol{R}'\}} U(\boldsymbol{R} - \boldsymbol{R}')\delta(y - R^y)\delta(y' - R'^y), \tag{B2}$$

where $\boldsymbol{R}$ and $\boldsymbol{R}'$ run over all atomic sites. The matrix element (14) specialized to (B2), after integration over coordinates $y$ and $y'$, is:

$$V_{n,m;s,t} = \frac{L_y^2}{4 N_c^2 \ell_{\text{QD}}^2} \sum_{\{\boldsymbol{R}\}} \sum_{\{\boldsymbol{R}'\}} F_n^*(R^y) F_m^*(R'^y)$$
$$\times\ U(\boldsymbol{R} - \boldsymbol{R}') F_s(R'^y) F_t(R^y). \tag{B3}$$

Going into the continuum limit by following the same procedure used to derive (A8) gives

$$V_{n,m;s,t} = \frac{1}{L_x^2 L_z^2 \ell_{\text{QD}}^2} \iint F_n^*(y) F_m^*(y')$$
$$\times\ U(\boldsymbol{r} - \boldsymbol{r}') F_s(y') F_t(y)\,d\boldsymbol{r}\,d\boldsymbol{r}'. \tag{B4}$$

To proceed it is convenient to write $U(\boldsymbol{r} - \boldsymbol{r}')$ [Eq. (16)] as

$$U_0\left\{1 + \mathcal{L}^{-2}\left[4R^2 \sin^2\left(\frac{x - x'}{2R}\right) + (y - y')^2 + a_z^2\right]\right\}^{-1/2}$$

where we have replaced $(z - z')^2$ with the constant value $a_z$, which is the average distance between a $2p_z$ electron and the nucleus of the carbon atom,[38] $a_z = 3a_B = 1.587$ Å. Hence, integrating over the coordinates $z$ and $z'$ cancels out the factor $1/L_z^2$ in (B4). By further integrating over $x$ and $x'$ and comparing the result with (14) allows for identifying $U_{\text{FW}}(y - y')$ as Eq. (17).

### 2. BW integrals

The two-body matrix element (14) specialized to (B1) with indices $(\tau, -\tau; \tau, -\tau)$ is

$$V_{n,m;s,t}(\tau) = \frac{L_y^2}{4N_c^2 \ell_{\mathrm{QD}}^2} \sum_{pp'} [2\delta_{p,p'} - 1] e^{i\tau \phi_{pp'}}$$
$$\times \sum_{\{\boldsymbol{R}_p\}} \sum_{\{\boldsymbol{R}'_{p'}\}} e^{i\tau(\boldsymbol{M}'-\boldsymbol{M})\cdot(\boldsymbol{R}_p - \boldsymbol{R}'_{p'})} U(\boldsymbol{R}_p - \boldsymbol{R}'_{p'})$$
$$\times F_n^*(R_p^y) F_m^*(R_{p'}^{\prime y}) F_s(R_{p'}^{\prime y}) F_t(R_p^y), \qquad (B5)$$

with $\phi_{AA} = \phi_{BB} = 0$ and $\phi_{AB} = -\phi_{BA} = 2\alpha + 5\pi/3$. We will eventually show that the matrix element (B5) does not depend on $\tau$. We may write $\boldsymbol{R}_p - \boldsymbol{R}'_{p'}$ appearing in (B5) as $\boldsymbol{R}_p - \boldsymbol{R}'_{p'} = \boldsymbol{R}_L + \boldsymbol{v}_{pp'}$, with $\boldsymbol{v}_{AA} = \boldsymbol{v}_{BB} = \boldsymbol{0}$ and $\boldsymbol{v}_{BA} = \boldsymbol{v} = -\boldsymbol{v}_{AB}$, where $\boldsymbol{R}_L$ is a lattice vector and $\boldsymbol{v}$ is the basis vector connecting B and A sites in graphene unit cell (cf. Fig. 1). Furthermore, the phase factor in (B5) may be written as

$$e^{i\tau(\boldsymbol{M}'-\boldsymbol{M})\cdot(\boldsymbol{R}_L + \boldsymbol{v}_{pp'})}$$
$$= e^{i\tau(\boldsymbol{K}'-\boldsymbol{K})\cdot(\boldsymbol{R}_L + \boldsymbol{v}_{pp'})} \cdot e^{+2i\nu\tau(\boldsymbol{R}_L + \boldsymbol{v}_{pp'})_x/(3R)}, \quad (B6)$$

where the first exponential is rapidly varying in real space in comparison with envelope functions since $\boldsymbol{K}' - \boldsymbol{K}$ is not a reciprocal lattice vector. Therefore, in the lattice site sums occurring in (B5) we keep only the leading terms (the shortest $\boldsymbol{R}_L$ close to the origin) and take $e^{+2i\nu\tau(\boldsymbol{R}_L + \boldsymbol{v}_{pp'})_x/(3R)} \approx 1$, obtaining:

$$V_{n,m;s,t}(\tau) \approx \frac{L_y^2}{4N_c^2 \ell_{\mathrm{QD}}^2} \sum_{pp'} [2\delta_{p,p'} - 1] e^{i\tau \phi_{pp'}}$$
$$\times \sum_{\{\boldsymbol{R}_L\}} e^{i\tau(\boldsymbol{K}'-\boldsymbol{K})\cdot(\boldsymbol{R}_L + \boldsymbol{v}_{pp'})} U(\boldsymbol{R}_L + \boldsymbol{v}_{pp'})$$
$$\times \sum_{\{\boldsymbol{R}_p\}} F_n^*(R_p^y) F_m^*(R_p^y) F_s(R_p^y) F_t(R_p^y). \qquad (B7)$$

We next focus on the quantity[21]

$$\tilde{U}_{\mathrm{BW}}^{pp'}(\tau) \equiv \sum_{\{\boldsymbol{R}_L\}} e^{i\tau(\boldsymbol{K}'-\boldsymbol{K})\cdot(\boldsymbol{R}_L + \boldsymbol{v}_{pp'})} U(\boldsymbol{R}_L + \boldsymbol{v}_{pp'}) \quad (B8)$$

appearing in (B7). As we show in Appendix B 3, $\tilde{U}_{\mathrm{BW}}^{AA}(\tau) = \tilde{U}_{\mathrm{BW}}^{BB}(\tau) = \tilde{U}_{\mathrm{BW}}$ (Ref. 21 provides $\tilde{U}_{\mathrm{BW}} = 4$ eV as an estimate), and $\tilde{U}_{\mathrm{BW}}^{pp'} = 0$ with $p \neq p'$ (hence $\tilde{U}_{\mathrm{BW}}^{pp'}$ does not depend on $\tau$ since the sums on $\boldsymbol{R}_L$ and $-\boldsymbol{R}_L$ are equivalent). Therefore, by going into the continuum limit, (B7) becomes

$$V_{n,m;s,t} = \frac{L_y}{2N_c \ell_{\mathrm{QD}}^2} \tilde{U}_{\mathrm{BW}} \int F_n^*(y) F_m^*(y) F_s(y) F_t(y) \, dy. \qquad (B9)$$

Since the CNT surface area is $2\pi R L_y = A_{\mathrm{graph}} N_c$, with $A_{\mathrm{graph}} = \sqrt{3}a^2/2$ being the area of graphene unit cell, (B9) may be written as

$$V_{n,m;s,t} = \frac{\sqrt{3}a^2}{8\pi R} \ell_{\mathrm{QD}}^{-2} \tilde{U}_{\mathrm{BW}} \int F_n^*(y) F_m^*(y) F_s(y) F_t(y) dy, \qquad (B10)$$

from which Eq. (18) immediately follows.

### 3. Properties of $\tilde{U}_{\mathrm{BW}}^{pp'}(\tau)$

The only non trivial property to be demonstrated is that $\tilde{U}_{\mathrm{BW}}^{AB} = 0$. It is convenient to write the definition (B8) in the original graphene reference frame (cf. Fig. 1). Then

$$\tilde{U}_{\mathrm{BW}}^{AB} = \sum_{\{\boldsymbol{R}_A\}} e^{i\tau(\boldsymbol{K}'-\boldsymbol{K})\cdot \boldsymbol{R}_A} U(\boldsymbol{R}_A). \qquad (B11)$$

The sublattice $\{\boldsymbol{R}_A\}$ is invariant under the rotation $\mathbb{R}^{-1}(\pm 2\pi/3)$ of all its elements around the origin. Since $U(\boldsymbol{r})$ is rotationally invariant, (B11) may be written as

$$\tilde{U}_{\mathrm{BW}}^{AB} = \frac{1}{3} \sum_{\{\boldsymbol{R}_A\}} U(\boldsymbol{R}_A) \left\{ e^{i\tau(\boldsymbol{K}'-\boldsymbol{K})\cdot \mathbb{R}^{-1}\left(+\frac{2\pi}{3}\right)\boldsymbol{R}_A} \right.$$
$$\left. + e^{i\tau(\boldsymbol{K}'-\boldsymbol{K})\cdot \boldsymbol{R}_A} + e^{i\tau(\boldsymbol{K}'-\boldsymbol{K})\cdot \mathbb{R}^{-1}\left(-\frac{2\pi}{3}\right)\boldsymbol{R}_A} \right\}. \qquad (B12)$$

In terms of integer indices locating $\boldsymbol{R}_A$, (B12) reads as

$$\tilde{U}_{\mathrm{BW}}^{AB} = \frac{1}{3} \sum_{n_1} \sum_{n_2} U(n_1, n_2) e^{i\tau 2\pi(n_1+n_2)/3}$$
$$\times \left\{ e^{-i\tau 2\pi/3} + 1 + e^{+i\tau 2\pi/3} \right\}. \qquad (B13)$$

The expression within the brackets in (B13) is zero. QED.

### Appendix C: Effect of spin-orbit interaction on WM energy levels

The spin-orbit interaction operator which appears in the many-body Hamiltonian $\hat{H}$ [Eq. (10)] has the form (with $\nu = +1$) $\hat{H}_{\mathrm{SO}} = \Delta_{\mathrm{SO}} \gamma \hat{\eta}/R$, with

$$\hat{\eta} = \sum_{n\sigma\tau} \sigma\tau \hat{c}_{n\sigma\tau}^\dagger \hat{c}_{n\sigma\tau} \qquad (C1)$$

being the total helicity operator. For $N$ electrons, there are $N+1$ distinct eigenvalues of $\hat{\eta}$,

$$\eta = -N, (-N+2), \ldots, (N-2), N.$$

Since $\hat{\eta}$ commutes with $\hat{H}$ (except for the $\hat{V}_{\mathrm{BW}}$ term, safely negligible in the WM limit), as well as with $\hat{S}_y$, $\hat{T}_y$, the eigenstates of $\hat{H}$ may be labeled by the sets ot quantum numbers $(S_y, T_y, \eta)$. Therefore, $\hat{H}_{\mathrm{SO}}$ splits each WM multiplet, which is $4^N$-fold degenerate when $B = 0$, $\Delta_{\text{SO}} = 0$, into $N + 1$ components (see Fig. 9). The lowest component has minimum helicity, $\eta = -N$. This is consistent with states fully polarized both in spin and isospin, $S_y = \pm N/2$ and $T_y = \mp N/2$. Furthermore, the magnetic field energetically favors one of the latter. Therefore, WM states are (iso)spin polarized without effort even in the presence of spin-orbit interaction, as confirmed by ED data discussed in Sec. VI.

TABLE II: Classical equilibrium positions of the WM.

| $N$ | $\{\bar{y}_i/\Lambda\}$ |
|---|---|
| 2 | $\{-1, +1\}$ |
| 3 | $\{-\sqrt[3]{5}, 0, +\sqrt[3]{5}\}$ |
| 4 | $\{-wW, -W, +W, +wW\}$ |
| 5 | $\{-\sqrt[3]{5}vV, -\sqrt[3]{5}V, 0, +\sqrt[3]{5}V, +\sqrt[3]{5}vV\}$ |

The residual degeneracies of the multiplet components may be easily worked out. To this aim, let us define the 'on-site' helicity $\eta(i) = \sigma(i)\tau(i)$ of the $i$th particle $(i = 1, \ldots, N)$, where here we consider distinguishable electrons localized at classical equilibrium positions $\bar{y}_i$. For the fundamental multiplet, $\eta(i) = -1\ \forall i$, with the value of $\tau(i)$ being uniquely determined once $\sigma(i)$ is fixed. If there are $k$ spin-down electrons, the number of possible combinations of $\sigma(i)$ and $\tau(i)$ giving $\eta = -N$ is $N!/[k!(N-k)!]$, and since $k$ may run from 0 to $N$ the degeneracy of the lowest multiplet is

$$\sum_{k=0}^{N} \frac{N!}{k!(N-k)!} = 2^N,$$

as reported in Fig. 9. For generic multiplets, $\eta = N - 2\ell$, with $\ell$ being the number of sites with $\eta(i) = -1$ ($\ell = 0, \ldots, N$). Using similar arguments, the degeneracy is found to be

$$2^\ell \cdot 2^{N-\ell} \frac{N!}{\ell!(N-\ell)!} = 2^N \frac{N!}{\ell!(N-\ell)!},$$

consistently with the data of Fig. 9. By summing over all $N + 1$ multiplet components, one of course recovers the initial degeneracy:

$$\sum_{\ell=0}^{N} 2^N \frac{N!}{\ell!(N-\ell)!} = 4^N.$$

### Appendix D: WM equilibrium positions and normal modes of vibration

Analytical expressions for the classical equilibrium positions $\bar{y}_i$ $(i = 1, \ldots, N)$ of the WM are provided in Table II, where the length $\Lambda$ is the equilibrium coordinate $\bar{y}_2$ for $N = 2$. Explicitly,

$$\Lambda = \left(\frac{e^2}{4\epsilon\omega_0^2 m^*}\right)^{1/3}.$$

Besides, $w$ is the real solution of

$$w^7 - 2w^5 - 25w^4 + w^3 - 6w^2 - 1 = 0,$$

$$W = \left[\frac{w^4 + 2w^3 + 10w^2 + 2w + 1}{w^2(w+1)^3}\right]^{1/3},$$

$$V = \left(\frac{13v^4 - 2v^2 + 5}{29v^4 - 2v^2 + 5}\right)^{1/3},$$

where $v$ is the real solution of

$$5v^7 - 10v^5 - 29v^4 + 5v^3 + 2v^2 - 5 = 0.$$

Numerically, one has $w = 3.162120$, $W = 0.721282$, $V = 0.763171$, $v = 2.120060$. These positions also hold for Table III, which reports the classical eigenfrequencies and eigenvectors of the WM. The normal coordinates $Y_i$ are linear combinations of the original coordinates $y_i$ through coefficients proportional to the components of the eigenvectors $(i = 1, \ldots, N)$.

---

TABLE III: Classical eigenfrequencies and eigenvectors of the WM. The constant $\beta$ takes the value $\beta = -0.521741$.

| $N$ | $\{\omega_i/\omega_0\}$ | numerical value | eigenvectors (a.u.) |
|---|---|---|---|
| 2 | 1 | 1.000 | $(+1\ +1)$ |
|   | $\sqrt{3}$ | 1.732 | $(+1\ -1)$ |
| 3 | 1 | 1.000 | $(+1\ +1\ +1)$ |
|   | $\sqrt{3}$ | 1.732 | $(+1\ 0\ -1)$ |
|   | $\sqrt{29/5}$ | 2.408 | $(+1\ -2\ +1)$ |
| 4 | 1 | 1.000 | $(+1\ +1\ +1\ +1)$ |
|   | $\sqrt{3}$ | 1.732 | $(+w\ +1\ -1\ -w)$ |
|   | $\left[\frac{w^5-41w^4-2w^3-118w^2+w-1}{w^5-9w^4-2w^3-22w^2+w-1}\right]^{1/2}$ | 2.410 | $(+1\ -1\ -1\ +1)$ |
|   | $\left[\frac{25w^4+10w^2-3}{3w^4-2w^2-1}\right]^{1/2}$ | 3.052 | $(-1\ +w\ -w\ +1)$ |
| 5 | 1 | 1.000 | $(+1\ +1\ +1\ +1\ +1)$ |
|   | $\sqrt{3}$ | 1.732 | $(+v\ +1\ 0\ -1\ -v)$ |
|   | $\left[\frac{53v^6-39v^4+79v^2-29-16\beta(6v^4-3v^2+1)}{13v^6-15v^4+7v^2-5}\right]^{1/2}$ | 2.412 | $(+1\ \beta\ -2-2\beta\ \beta\ +1)$ |
|   | $\left[\frac{87v^6+19v^4+37v^2-15}{13v^6-15v^4+7v^2-5}\right]^{1/2}$ | 3.055 | $(-1\ +v\ 0\ -v\ +1)$ |
|   | $\left[\frac{(841v^6-579v^4+203v^2)/5-29+16\beta(6v^4-3v^2+1)}{13v^6-15v^4+7v^2-5}\right]^{1/2}$ | 3.671 | $\left(+1\ \frac{-3-2\beta}{2+3\beta}\ \frac{2-2\beta}{2+3\beta}\ \frac{-3-2\beta}{2+3\beta}\ +1\right)$ |